\documentclass{aa}  

\usepackage{graphicx}
\usepackage{multirow}
\usepackage{txfonts}
\begin{document}

   \title{Nature of shocks revealed by SOFIA OI observations\\ in the Cepheus E protostellar outflow\thanks{This article uses \textit{Herschel}-PACS data; \textit{Herschel} is an ESA space observatory with science instruments provided by European-led Principal Investigator consortia and with important participation from NASA.}}
   \author{A. Gusdorf
          \inst{1}
          \and
          S. Anderl\inst{2,3}
          \and 
          B. Lefloch\inst{2,3}
          \and
          S. Leurini\inst{4, 5}
           \and
          H. Wiesemeyer\inst{4}
	\and
          R. G\"usten\inst{4}
          \and	
          M. Benedettini\inst{6}
          \and
          C. Codella\inst{7}
          \and
          \\B. Godard\inst{1}
          \and
          A. I. G\'omez-Ruiz\inst{8}
         \and
         K. Jacobs\inst{9}
	\and
	L. E. Kristensen\inst{10}
          \and 
	P. Lesaffre\inst{1}
         \and
	G. Pineau des For\^ets\inst{11, 1}
         \and 
         D. C. Lis\inst{1, 12}
           }

  	   \institute{LERMA, Observatoire de Paris, \'Ecole normale sup\'erieure, PSL Research University, CNRS, Sorbonne Universit\'es, UPMC Univ. Paris 06, F-75231, Paris, France; \email{antoine.gusdorf@lra.ens.fr}
	   \and
	   Univ. Grenoble Alpes, CNRS, IPAG, F-38000 Grenoble, France
	   \and
	   CNRS, IPAG, F-38000 Grenoble, France
             \and
             Max Planck Institut f\"ur Radioastronomie, Auf dem H\"ugel, 69, 53121 Bonn, Germany
             \and
             INAF-Osservatorio Astronomico di Cagliari, Via della Scienza 5, I-09047, Selargius (CA)
	   \and
             INAF - Istituto di Astrofisica e Planetologia Spaziali, Via del Fosso del Cavaliere 100, I-00133 Roma, Italy
	   \and
             INAF, Osservatorio Astrofisico di Arcetri, Largo Enrico Fermi 5, I-50125 Firenze, Italy
             \and
	   Instituto Nacional de Astrof\'isica, \'Optica y Electr\'onica, Luis E. Erro 1, 72840, Tonantzitla, Puebla, M\'exico
	   \and
	   I. Physikalisches Institut der Universit\"at zu K\"oln, Z\"ulpicher Strasse 77, 50937 K\"oln, Germany
	   \and
	   Centre for Star and Planet Formation, Niels Bohr Institute and Natural History Museum of Denmark, University of Copenhagen, \O ster Voldgade 5-7, DK-1350 Copenhagen K, Denmark
             \and
             Institut d'Astrophysique Spatiale, CNRS UMR 8617, Universit\'e Paris-Sud, 91405, Orsay, France
	   \and
	   California Institute of Technology, Cahill Center for Astronomy and Astrophysics 301-17, Pasadena, CA 91125, USA}


 
  \abstract
   {Protostellar jets and outflows are key features of the star-formation process, and primary processes of the feedback of young stars on the interstellar medium. Understanding the underlying shocks is necessary to explain how jet and outflow systems are launched, and to quantify their chemical and energetic impacts on the surrounding medium.}
   {We performed a high-spectral resolution study of the [OI]$_{\rm 63 \mu m}$ emission in the outflow of the intermediate-mass Class 0 protostar Cep E-mm. The goal is to determine the structure of the outflow, to constrain the chemical conditions in the various components, and to understand the nature of the underlying shocks, thus probing the origin of the mass-loss phenomenon.}
  {We present observations of the \ion{O}{I} $^3$P$_1 \rightarrow$ $^3$P$_2$, OH between $^2\Pi_{1/2}$ $J = 3/2$ and $J = 1/2$ at 1837.8~GHz, and CO (16--15) lines with the GREAT receiver onboard SOFIA towards three positions in the Cep E protostellar outflow: Cep E-mm (the driving protostar), Cep E-BI (in the southern lobe), and Cep E-BII (the terminal position in the southern lobe).}
   {The CO (16--15) line is detected at all three positions. The [OI]$_{\rm 63 \mu m}$ line is detected in Cep E-BI and BII, whereas the OH line is not detected. In Cep E-BII, we identify three kinematical components in \ion{O}{I} and CO. These were already detected in CO transitions and relate to spatial components: the jet, the HH377 terminal bow-shock, and the outflow cavity. We calculate line temperature and line integrated intensity ratios for all components. The \ion{O}{I} column density is higher in the outflow cavity than in the jet, which itself is higher than in the terminal shock. The terminal shock is the region where the abundance ratio of \ion{O}{I} to CO is the lowest (about 0.2), whereas the jet component is atomic ($N$(\ion{O}{I}) / $N$(CO)$\sim$2.7). In the jet, we compare the [OI]$_{\rm 63 \mu m}$ observations with shock models that successfully fit the integrated intensity of 10 CO lines. We find that these models most likely do not fit the [OI]$_{\rm 63 \mu m}$ data.}
   {The high intensity of \ion{O}{I} emission points towards the propagation of additional dissociative or alternative FUV-irradiated shocks, where the illumination comes from the shock itself. A picture emerges from the sample of low-to-high mass protostellar outflows, where similar observations have been performed, with the effects of illumination increasing with the mass of the protostar. These findings need confirmation with more observational constraints and a larger sample.}


   \maketitle
   
\titlerunning{SOFIA observations of shocked OI in the Cep E outflow}
\authorrunning{A. Gusdorf, S. Anderl, B. Lefloch, et al.}
%

\section{Introduction}
\label{sec:intro}

   \begin{figure*}
   \centering
   \includegraphics[width=15cm]{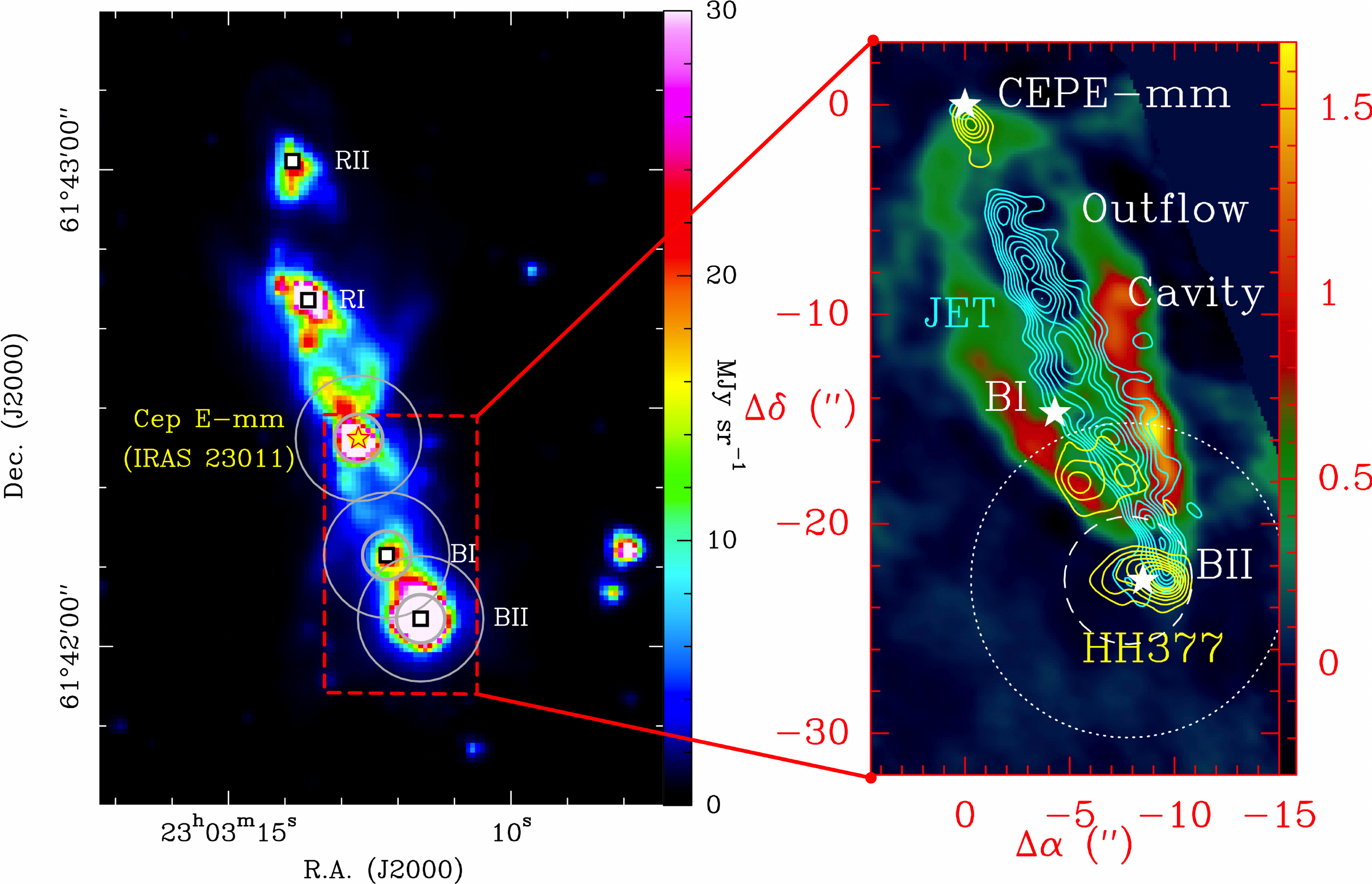}
      \caption{\textit{Left:} a finder's chart for the positions of interest in the Cep E protostellar outflow: \textit{Spitzer}-IRAC band-two (4.5 $\mu$m) image, retrieved from the \textit{Spitzer} archive. Grey circles mark the diffraction-limited SOFIA beams at the frequencies of [OI]$_{\rm 63 \mu m}$ (smaller circles) and CO (16--15) or OH at 1838~GHz (larger circles) lines. The three observed positions are indicated (Cep E-mm, red star, BI and BII, white rectangles), as well as others in the red lobe (RI and RII). The knots denomination is from \citet{Gomezruiz12}. \textit{Right:} zoom on the southern lobe of the outflow showing the spatial components identified by L15 in CO (2--1) (see Section~\ref{sec:results}): the jet (cyan contours), the terminal bow-shock (yellow contours distant from the protostar) and the cavity walls (coloured background). The three observed positions are indicated with a white star, and the SOFIA beams at the frequencies of [OI]$_{\rm 63 \mu m}$ (smaller circles) and CO (16--15) or OH at 1838~GHz (larger circles) lines are shown in BII.}
         \label{figure1}
   \end{figure*}

Protostellar outflows constitute the most prominent signposts of star formation. In the earliest phases, when the young protostar builds up most of its final mass by accreting material from its parental cloud, the ejection of magneto-centrifugal jets is believed to help the protostellar system to release angular momentum from its inner disk system. Slower wide-angle winds surrounding the fast jets stir up the ambient cloud environment and create molecular outflow cavities (e.g. \citealt{Arce07, Frank14}). These protostellar feedback processes play a major role in protostellar mass accretion, and in the interaction of the protostar with its parental environment. Their understanding is still subject to many open questions, for instance about the nature and universality of jets, their temporal activity, and their properties in terms of energy and momentum transfer. It is therefore of paramount importance to study the properties of protostellar jets in various protostellar sources, as has been done recently towards L1157 (e.g. \citealt{Podio16,Tafalla15}) and BHR71 (e.g. \citealt{Benedettini17}), for example.

Jets have been indirectly traced by bow-shock emission (or working surfaces, e.g. \citealt{Gueth98}), outflow cavities (or swept-up material, \citealt{Gueth96,Bachiller01}), and compact dissociative (e.g. \citealt{Kristensen13,Nisini15}) or high-velocity molecular (e.g. \citealt{Tafalla15}) shocks. However, a direct probe of jets in outflows requires observations of typical tracers that are spatially or spectroscopically resolved. One such typical jet tracer is the fine-structure line of atomic oxygen at 63 $\mu$m, which is one of the main coolants in jets. The German Receiver for Astronomy at TeraHertz frequencies (GREAT) in the Stratospheric Observatory for Infrared Astronomy (SOFIA) now allows spectroscopically resolved observations of [OI]$_{\rm 63 \mu m}$ and thereby offers this transition as a valuable tool for studying protostellar jet systems (for a detailed study on the potential of this tracer in the context of star formation we refer to \citealt{Leurini15}).

In order to fully exploit the potential of [OI]$_{\rm 63 \mu m}$ as a diagnostic tool, its interpretation relies on the use of detailed shock modelling. The Paris-Durham shock model (\citealt{Flower15}) describes slow ($\varv_s$ $\lesssim$ 50 km s$^{-1}$), $C$-, $J$-, and time-dependent magnetohydrodynamic (MHD) shocks, propagating in dark molecular environments. It internally computes the [OI]$_{\rm 63 \mu m}$ emissivities in optically thin local thermodynamical equilibrium (LTE) conditions. In faster shocks, the high temperature reached in the shock front triggers UV emission from the gas. This UV emission can propagate to the pre-shock medium and cause a radiative precursor that leads to photodissociation and ionization in that gas \citep{Hollenbach89}. For such faster shocks, with velocities up to 150 km s$^{-1}$, these authors provide [OI]$_{\rm 63 \mu m}$ emissivities for a range of pre-shock densities. A combination of these different models can be used to constrain the physical and dynamical characteristics of jet-driven shocks, as well as the conditions in the pre-shock environment by comparing model predictions with the observed intensities of [OI]$_{\rm 63 \mu m}$ and other available tracers.

In the present study, we applied this strategy to understand the nature of shocks in the Cep E protostellar outflow. This source is an intermediate-mass Class 0 protostar ($L = 100~L_\odot$, \citealt{Lefloch96, Chini01}) in the Cepheus OB3 association at a distance of 730 pc (\citealt{Sargent77}). Its prominent and well-studied outflow is driven by a jet of a size of 1.7$''$ $\times$ 21$''$, as identified in several CO transitions (\citealt{Gomezruiz12}) up to CO (16-15) and interpreted by means of time-dependent MHD shock models by \citet{Lefloch15} (hereafter L15). In particular, the CO emission of shocked gas along the jet was found to be consistent with very young (220--740~yr) shocks with velocities of 20--30 km s$^{-1}$, propagating into a pre-shock medium of density $n_{\rm H}$ = (0.5--1.0) $\times$ 10$^5$ cm$^{-3}$. However, this study already demonstrated the need for additional observations in order to achieve a more precise determination of the jet shock parameters. In this work, we aim at further constraining the nature of this protostellar jet using [OI]$_{\rm 63 \mu m}$ observations, together with an extended grid of MHD Paris-Durham shock models. This article is structured as follows: observations are introduced in Section~\ref{sec:obs} and direct results are presented in Section~\ref{sec:results}. In Section~\ref{sec:ab}, we infer abundances from the kinematical components identified in our observations. We compare our observations to sophisticated models of interstellar shocks in Section~\ref{sec:nature} to understand the nature of such shocks, while Section~\ref{sec:conc} contains our concluding remarks.

\section{Observations}
\label{sec:obs}

\begin{table}[h]
\caption{Observed positions and sensitivity (r.m.s. of baseline noise) at 1~km~s$^{-1}$ spectral resolution for each observed line.}            
\label{table1}      
\centering                          
\begin{tabular}{l  c  c  c}        
\hline        
\hline              
position & Cep E-mm & Cep E-BI & Cep -BII \\
\hline
$\alpha$ (J2000) & 23$^{\rm h}$03$^{\rm m}$12$\fs$7 & 23$^{\rm h}$03$^{\rm m}$12$\fs$2 & 23$^{\rm h}$03$^{\rm m}$11$\fs$6\\
$\delta$ (J2000) & 61$^{\circ}$42$'$26$\farcs$2 & 61$^{\circ}$42$'$11$\farcs$5 & 61$^{\circ}$42$'$03$\farcs$5\\
\hline
r.m.s. OH (K) & 0.19 & 0.05 & 0.06 \\
r.m.s. CO (K) & 0.21 & 0.12 & 0.06 \\
r.m.s. \ion{O}{I} (K) & 0.19 & 0.04 & 0.07 \\
\hline           
\hline           
\end{tabular}
\end{table}

\begin{table*}
\caption{Characteristics and observational parameters of the OH transitions between the $^2\Pi_{1/2}$ $J = 3/2$ and $J = 1/2$ states, and of the [OI]$_{\rm 63 \mu m}$ and CO (16--15) lines. $A(B)\equiv A \times 10^B$. The `shift' column contains the velocity shift relative to the component with the largest Einstein $A$ coefficient for he OH transition. Source: JPL \citep{Pickett98}.}            
\label{table2}      
\centering                          
\begin{tabular}{l  c  c  c  c  c  c c c c c c c}        
\hline           
\hline
\tiny{triplet} & \tiny{transition} & \tiny{frequency} & \tiny{$A_{\rm{ul}}$} & \tiny{$g_{\rm u}$} & \tiny{$g_{\rm l}$} & \tiny{$E_{\rm u}$} & \tiny{shift} & \tiny{beam size} & \tiny{spectral resolution} & \tiny{beam} & \tiny{forward} & \tiny{$T_{\rm sys}$}  \\
\tiny{properties} & \tiny{$F^{\prime}_{p^{\prime}} \rightarrow F_p$} & \tiny{(GHz)} & \tiny{(s$^{-1}$)} & & & \tiny{(K)} & \tiny{(km s$^{-1}$)} & \tiny{($''$)} & \tiny{(km~s$^{-1}$)} & \tiny{efficiency} & \tiny{efficiency} & \tiny{(K)}  \\
\hline
\hline
\tiny{OH} & \tiny{$1+ \rightarrow 1-$} & \tiny{1837.7466} & \tiny{2.1(-2)} & \tiny{3} & \tiny{3} & \tiny{270.1} & \tiny{11.5} & & & & &  \\
\tiny{1838~GHz} & \tiny{$2+ \rightarrow 1-$} & \tiny{1837.8168} & \tiny{6.4(-2)} & \tiny{5} & \tiny{3} & \tiny{270.1} & \tiny{0.0} & \tiny{15.3} & \tiny{1.00} & \tiny{0.67} & \tiny{0.97} & \tiny{4930 - 5324}  \\
\tiny{163.1~$\mu$m} & \tiny{$1+ \rightarrow 0-$} & \tiny{1837.8370} & \tiny{4.3(-2)} & \tiny{3} & \tiny{1} & \tiny{270.1} & \tiny{-3.3} & & & & &    \\
\hline
\tiny{CO} & \tiny{(16--15)} & \tiny{1841.3455} & \tiny{4.05(-4)} & \tiny{33} & \tiny{31}  & \tiny{751.7} & & \tiny{15.3} & \tiny{0.99} & \tiny{0.65} & \tiny{0.97} & \tiny{4930 - 5324} \\
\hline
\tiny{\ion{O}{I}} & \tiny{$^3$P$_1 \rightarrow$ $^3$P$_2$} & \tiny{4744.7775} & \tiny{8.91(-5)} & \tiny{3} & \tiny{5} & \tiny{227.7} & & \tiny{6.1} & \tiny{0.99} & \tiny{0.67} & \tiny{0.97} & \tiny{2997 - 4232} \\
\hline
\hline
\end{tabular}
\end{table*}

The observations of the Cep E protostellar outflow were conducted with the GREAT\footnote{GREAT is a development by the MPI f\"ur Radioastronomie and the KOSMA$/$Universit\"at zu K\"oln, in cooperation with the MPI f\"ur Sonnensystemforschung and the DLR Institut f\"ur Planetenforschung.} receiver \citep{Heyminck12} during three SOFIA flights on 15, 16, and 18 December 2015 (legs 12, 8, and 8, respectively ), as part of the Cycle 3 Community science program. Three positions were observed: Cep E mm, BI, and BII (see Fig.~\ref{figure1} and Table~\ref{table1}). 

The observation of the [OI]$_{\rm 63 \mu m}$ line is possible in the H channel of GREAT using a hot electron bolometer heterodyne mixer \citep{Buechel15}, with a 2.5~GHz bandwidth receiver and a spectral resolution of 0.99~km~s$^{-1}$. For the tuning to the [OI]$_{\rm 63 \mu m}$ line, a quantum-cascade laser was used as local oscillator \citep{Richter15}. At the Cep E-mm position, the zero-level width observed for example by \citet{Lefloch11} in SiO or H$_2$O lines is wider than 200~km~s$^{-1}$. This is more than the bandwidth available at the 4744.778~GHz frequency of the [OI]$_{\rm 63 \mu m}$ line ($\sim$~160~km~s$^{-1}$). For this reason, we covered the emission of this line in two spectral windows. For the BI and BII positions, we prepared our tuning setup based on the CO line profiles showed by L15 at the BII position, which were about 150~km~s$^{-1}$ wide (at baseline level). A single USB tuning was sufficient to cover the [-150;10]~km~s$^{-1}$ interval at the frequency of the [OI]$_{\rm 63 \mu m}$ line. 

For OH and CO observations of all three positions, the lower frequency, L2 channel was connected to 4~GHz wide digital back-ends described in \citet{Risacher16}, providing respective spectral resolutions of 1.00 and 0.99 km s$^{-1}$ (given in Table~\ref{table2}). This channel was tuned in LSB to the frequency of the OH triplet between $^2\Pi_{1/2}$ $J = 3/2$ and $J = 1/2$ at 1837.817~GHz. This setup allows us to pick up the CO (16--15) in the upper sideband at 1841.346 GHz (see Table~\ref{table2} for spectroscopic parameters). 

The resulting r.m.s. uncertainties are given in Table~\ref{table1}. The observations were performed in double beam-switching mode with an amplitude of 40$''$ (80$''$ throw) at the position angle of 45$^\circ$ and a phase time of 0.5 sec. The nominal focus position was updated regularly against temperature drifts of the telescope structure. The pointing was established with the optical guide cameras to an accuracy of better than $\sim$1$''$. The beam widths and efficiencies are indicated in Table~\ref{table2}. The data were calibrated with the KOSMA/GREAT calibrator \citep{Guan12}, which removes residual telluric lines, and were further processed (mostly by removing linear baseline) with the CLASS software\footnote{http://www.iram.fr/IRAMFR/GILDAS}. 

\section{Results}
\label{sec:results}

\subsection{Line profiles}
\label{sub:lp}

   \begin{figure}
   \centering
   \includegraphics[width=9cm]{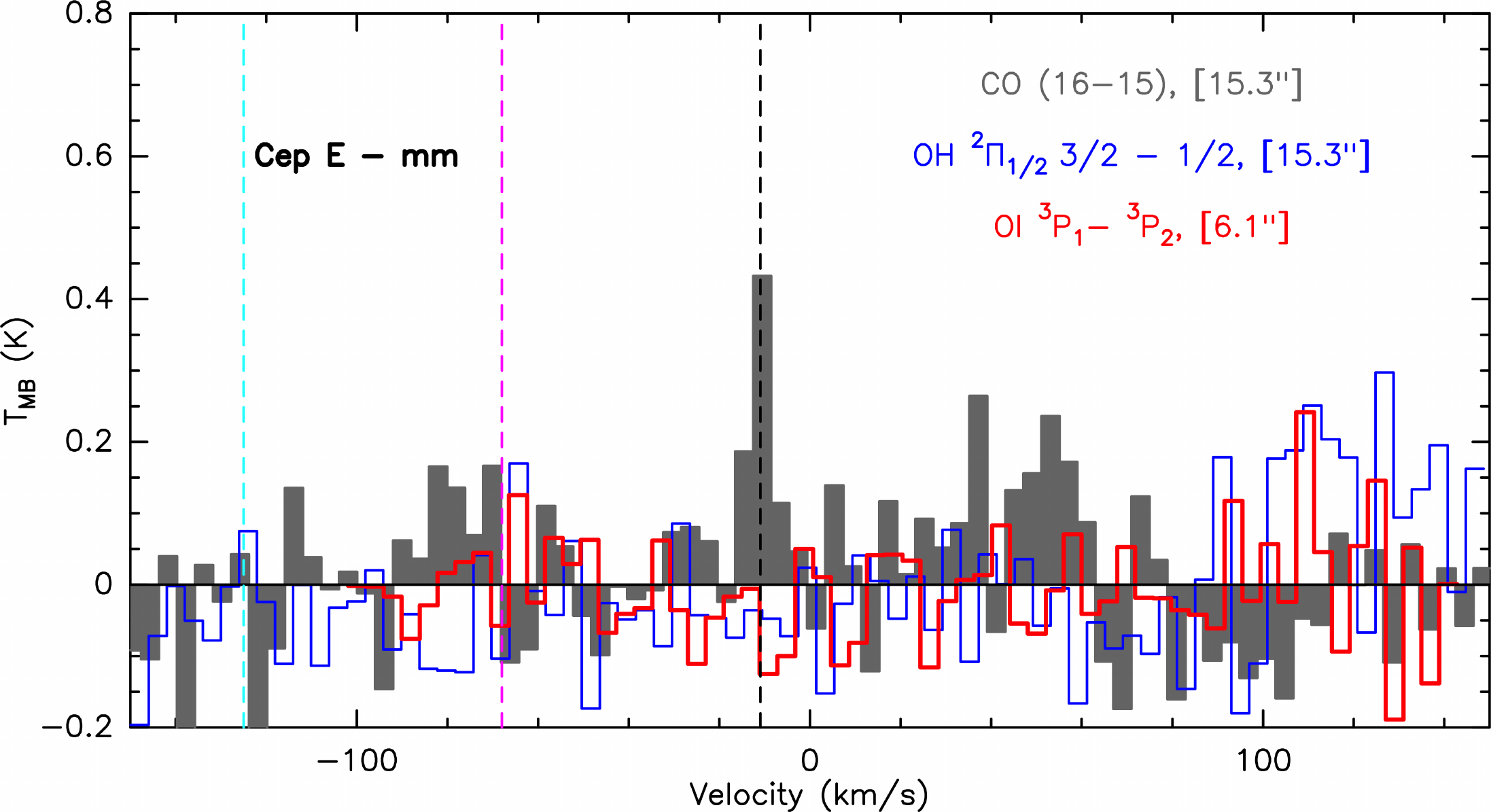}
   \includegraphics[width=9cm]{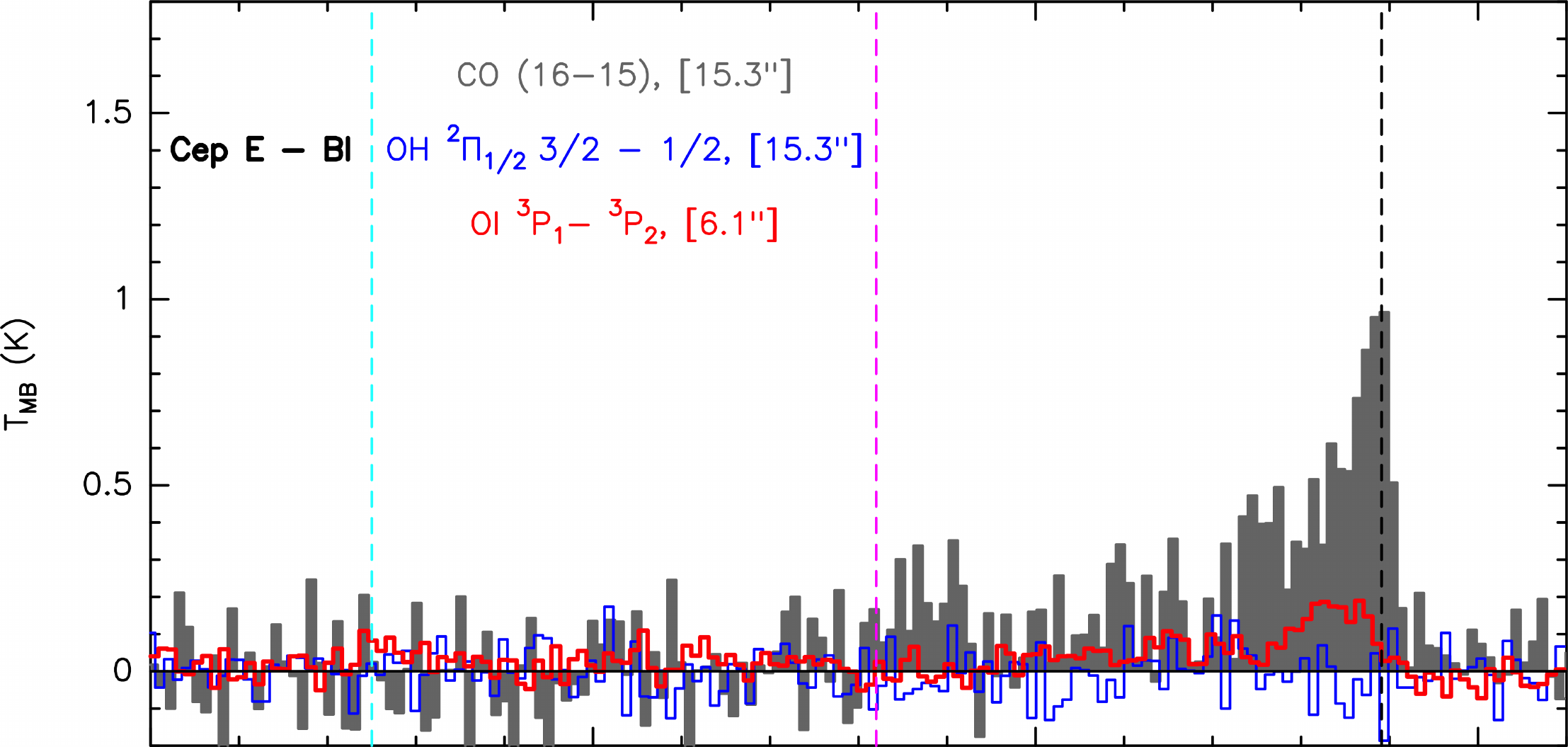}
   \includegraphics[width=9cm]{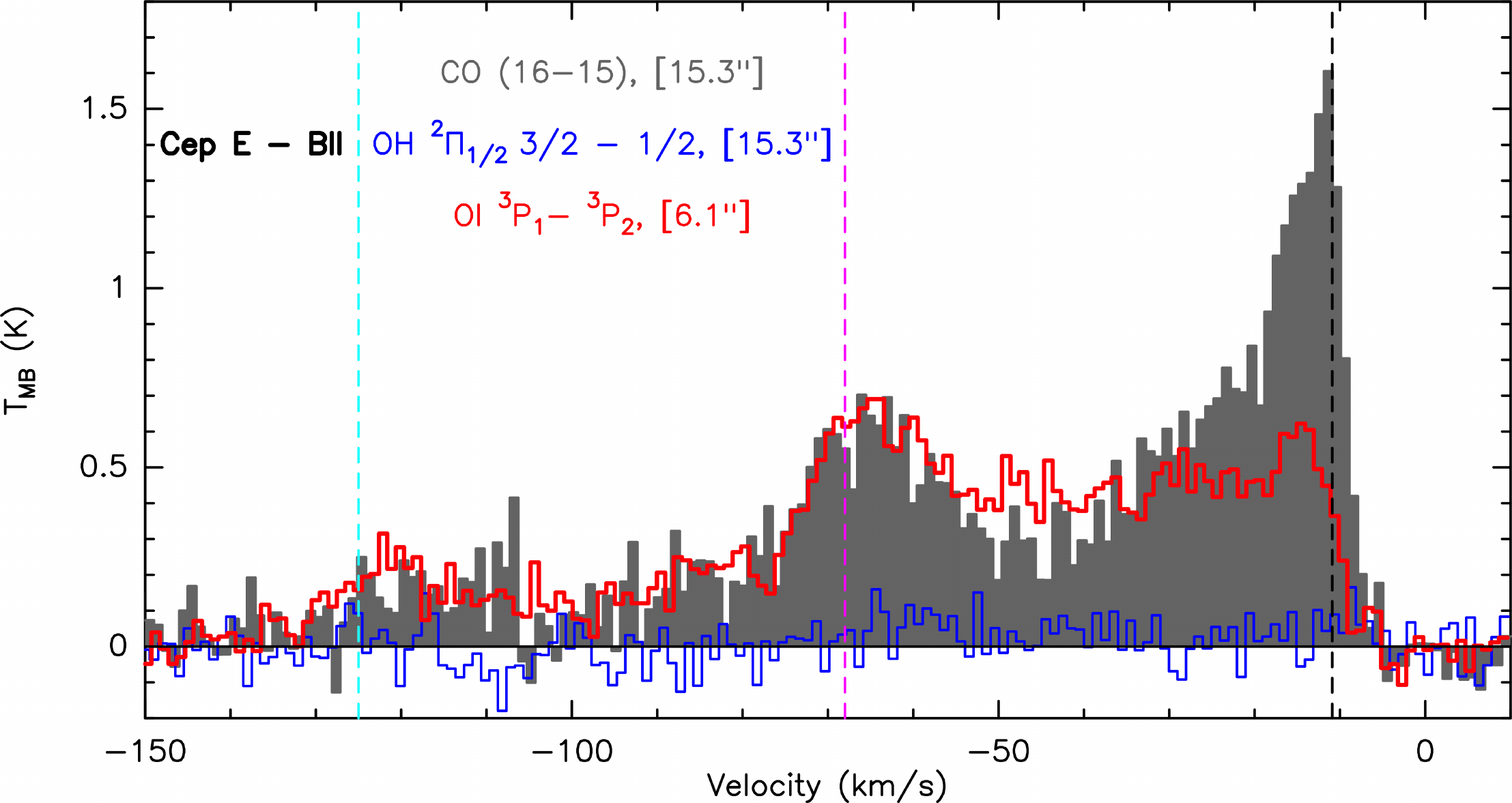}
      \caption{CO (16--15) (grey histograms), [OI]$_{\rm 63 \mu m}$ (red line), and OH at 1838~GHz (blue line) observations obtained in Cep E-mm (top panel), Cep E-BI (middle panel), and Cep E-BII (lower panel). The vertical dashed lines represent the central value for the velocity components identified by L15 in CO transitions: $-$125 (cyan), $-$68 (pink), and $-$10.9~km~s$^{-1}$ (black; systemic velocity of Cep E). The spectral resolution is 1~km~s$^{-1}$ for the Cep E-BI and BII spectra, and 4~km~s$^{-1}$ for Cep E-mm.}
         \label{figure2}
   \end{figure}

The spectra of the CO (16--15), [OI]$_{\rm 63 \mu m}$, and OH 1838~GHz lines obtained at the three positions are shown in Fig.~\ref{figure2}. L15 mapped the Cep E protostellar outflow in CO (2--1) at high angular resolution with the Plateau de Bure Interferometer (PdBI) and performed pointed CO observations with numerous single-dish telescopes at the Cep E-BII position. This allowed them to spectrally and spatially identify three components in the outflow, which we refer to throughout this article: 
\begin{itemize}
\item a narrow jet emitting between $-$140 and $-$110, centred on $-$125~km~s$^{-1}$, with a typical line width of 15~km~s$^{-1}$, well separated from the bipolar outflow wing emission;
\item the outflow cavity, composed of the gas contained in the cavity walls, that is, of the cavity walls plus the entrained gas that is not part of the jet that is emitted from the ambient cloud velocity ($-$10.9~km~s$^{-1}$, \citealt{Lefloch96}) up to $-$100~km~s$^{-1}$;
\item the terminal bow-shock labelled \lq HH377', which is emitted between $-$90 and $-$50~km~s$^{-1}$ and appears as a bump between the outflow cavity and the jet in the line profile, and whose CO (16--15) intensity is similar to the entrained gas at $-$20~km~s$^{-1}$.
\end{itemize}

At the protostar position, all lines are undetected except for a faint 3$\sigma$ emission of the CO (16--15) line around the systemic velocity of Cep E (at $-$10.9~km~s$^{-1}$). Hints of emission can be seen in the blue- and red-shifted wing, between $-$90 and $-$60, and 40 and 60~km~s$^{-1}$, respectively.

At the Cep E-BI and BII positions, only OH is undetected. At the Cep E-BI position, only the spectral component that was found by L15 to be associated with the walls of the outflow cavity is detected in CO and \ion{O}{I}. In CO, this component exhibits a wing that extends up to $-$75~km~s$^{-1}$, whereas the corresponding \ion{O}{I} emission is found to be much narrower. 

Finally, at the Cep E-BII position, the three components identified by L15, that is, the jet, the terminal bow-shock, and the outflow cavity, appear in the CO (16-15) and in [OI]$_{\rm 63 \mu m}$ line profiles. At the BI and BII positions, a slight shift in velocity of the order of 5~km~s$^{-1}$ might exist between the peak of the CO (16--15) emission and that of [OI]$_{\rm 63 \mu m}$. This shift could be a consequence of the two lines tracing a different material, or it could be an effect of self-absorption of [OI]$_{\rm 63 \mu m}$ near the systemic velocity. In particular, the bright bump in the entrained gas seen at $-$20~km~s$^{-1}$ in CO (16--15) appears between $-$30 and $-$20~km~s$^{-1}$ in [OI]$_{\rm 63 \mu m}$. Additionally, this bright bump is weaker than the HH377 component in [OI]$_{\rm 63 \mu m}$, in contrast to the CO (16--15) case. This could be an effect of different filling factors resulting from the different telescope beams at 4744.778~GHz and 1841.3455~GHz (se Fig.~\ref{figure1}). As there is no [OI]$_{\rm 63 \mu m}$ detection at the protostar position, and as the BI and BII observations of CO (16--15) are not independent (see Fig.~\ref{figure1}), we focused on the spectra obtained at the BII position.

\subsection{Line ratios}
\label{sub:lr}

We extracted integrated intensities from the [OI]$_{\rm 63 \mu m}$ and CO (16--15) spectra at the BII position. We based our analysis on the three kinematical components identified by L15 (see previous section): the jet at high velocities, the terminal bow-shock at intermediate velocities, and the outflow cavity at lower velocities. Our method was the same for the [OI]$_{\rm 63 \mu m}$ and CO (16--15) lines, given the similarity of their profiles. The method that we used to extract the integrated intensity from each kinematical component is fully consistent with that of L15:
\begin{itemize}
\item for the jet component, we integrated the emission between $-$140 and $-$100~km~s$^{-1}$;
\item for the HH377 component, we considered the velocity interval between $-$100 and $-$25~km~s$^{-1}$, in which we removed the outflow cavity component by fitting a first-order baseline, and then integrated the emission of the residual between $-$90 and $-$40~km~s$^{-1}$;
\item for the outflow cavity, we integrated the emission between $-$100 and $-$14~km~s$^{-1}$, from which we subtracted the integrated intensity associated with the HH377 terminal shock.
\end{itemize}
The final values that we extracted from this analysis can be found in Table~\ref{table3}. We used conservative error bars of $\pm$10\%, as the 3$\sigma$ value probably does not reflect the uncertainty intrinsic to the extraction we performed. 

\begin{table}
\tiny
\caption{
Raw integrated intensities extracted from the [OI]$_{\rm 63 \mu m}$ and CO (16--15) spectra, in units of K~km~s$^{-1}$, and corresponding line ratio, kinematic component by kinematic component (jet, bow-shock, outflow cavity).} 
\label{table3}      
\centering                          
\begin{tabular}{l c c c}     
\hline\hline           
component    &   jet & bow-shock & outflow cavity \\

\hline\hline

$\int T_{\rm mb}\, d\varv$  OI $^3$P$_1 \rightarrow$ $^3$P$_2$ & 5.3$\pm$0.5 & 5.8$\pm$0.6 & 28.1$\pm$2.8 \\
(K km s$^{-1}$) &  & & \\
\hline
$\int T_{\rm mb}\, d\varv$ CO (16--15) & 4.5$\pm$0.5 & 6.0$\pm$0.6 & 30.9$\pm$3.1 \\
(K km s$^{-1}$) & & &  \\
\hline
ratio [OI]/CO & 1.2$\pm$0.2 & 1.0$\pm$0.1 & 0.9$\pm$0.1\\
\hline\hline 
\end{tabular}
\end{table}

The table indicates similar integrated intensities for the jet and the terminal bow-shock components in [OI]$_{\rm 63 \mu m}$ and CO (16--15). These integrated intensities are significantly lower than the contribution from the outflow cavity. The line ratios found for the three components are similar.

\section{Abundances}
\label{sec:ab}

\subsection{Filling factors}
\label{sub:fitting}

We then estimated filling factor corrections to the integrated intensities. Based on their CO (2--1) interferometric observations, L15 inferred the size of the emission region for the jet, bow-shock, and outflow cavity components: 1.7$'' \times$21$''$, 4.5$''$, and 22$'' \times$10$''$. The authors were then able to infer filling factors for the jet, shock, and cavity wall components of their SOFIA/GREAT observations of CO (16--15) based on the assumption that the emission region was the same. As our observations were pointed in exactly the same BII position, we used the filling factor values derived from their observations for CO (16--15): 0.09 for the jet, 0.06 for the shock, and 0.15 for the cavity walls (see Table~\ref{table4}).

For the [OI]$_{\rm 63 \mu m}$ line emission, our method relies on the combined use of the \textit{Herschel}-PACS footprint (see Section~\ref{sec:PACS} of the Appendix and Fig.~\ref{figurea1}) obtained at this wavelength, of the \textit{Spitzer}-IRAC band-two (4.5 $\mu$m) image (see Fig.~\ref{figure1}), and of the CO (2--1) interferometric map of L15 (also see Fig.~\ref{figure1}):
\begin{itemize}
\item the outflow cavity most likely dominates the emission in the PACS footprint. In this map, the emission region has a typical deconvolved size of 10$''$. Given that the map is not Nyquist-sampled, and given the significantly different beam sizes for CO (16--15) and [OI]$_{\rm 63 \mu m}$ lines, we adopted a simple filling factor of 1 for this component in [OI]$_{\rm 63 \mu m}$;
\item the size of HH377 is measured in CO (2--1) ($\sim$4.5$''$, L15) and in \textit{Spitzer}-IRAC band-two emission (deconvolved size of about 10$''$). As HH377 emission is visible in the low-excitation CO (2--1) and the high-excitation CO (16--15) line, we consider that its typical size in [OI]$_{\rm 63 \mu m}$ is between 4.5 and 10$''$, yielding a filling factor of 0.55$\pm$0.15;
\item finally, we inferred the filling factor for the jet emission from the interferometric CO (2-1) map because this component is similar in CO (2--1) and  [OI]$_{\rm 63 \mu m}$: 0.25$\pm$0.05. 
\end{itemize} 

\begin{table}
\tiny
\caption{Filling-factor-corrected, integrated intensities extracted from the [OI]$_{\rm 63 \mu m}$ and CO (16--15) spectra, in units of K~km~s$^{-1}$ and erg~cm$^{-2}$~s$^{-1}$~sr$^{-1}$, and corresponding line ratio, kinematic component by kinematic component (jet, bow-shock, outflow cavity).}            
\label{table4}      
\centering                          
\begin{tabular}{l c c c}     
\hline\hline           
component    &   jet & bow-shock & outflow cavity \\

\hline\hline
filling factor OI $^3$P$_1 \rightarrow$ $^3$P$_2$ & 0.25$\pm$0.05 & 0.55$\pm$0.15 & <1 \\
filling factor CO (16--15) & 0.09 & 0.06 & 0.15\\

\hline

$\int T_{\rm mb}\, d\varv$  OI $^3$P$_1 \rightarrow$ $^3$P$_2$ & 21.2$\pm$4.7 & 10.5$\pm$3.1 & >28.1 \\
(K km s$^{-1}$) &  & &  \\

$\int T_{\rm mb}\, d\varv$ CO (16--15) & 50.0$\pm$5 & 100.0$\pm$10 & 206.0$\pm$20.6 \\
(K km s$^{-1}$) &  & &   \\

ratio [OI]/CO & 0.4$\pm$0.1 & 0.11$\pm$0.03 & >0.14\\

\hline

$\Sigma I_{\rm \nu} \Delta \nu$  OI $^3$P$_1 \rightarrow$ $^3$P$_2$ & 23.2$\pm$5.1 & 11.5$\pm$3.4 & > 30.7 \\
(10$^{-4}$ erg cm$^{-2}$ s$^{-1}$ sr$^{-1}$) &  &  & \\

$\Sigma I_{\rm \nu} \Delta \nu$  CO (16--15) & 3.2$\pm$0.3 & 6.4$\pm$0.6 & 13.2$\pm$1.3 \\
(10$^{-4}$ erg cm$^{-2}$ s$^{-1}$ sr$^{-1}$) & &  &  \\

ratio [OI]/CO & 7.3$\pm$1.8 & 1.8$\pm$0.6 & > 2.3\\

\hline\hline
\end{tabular}
\end{table}

The filling factors are indicated in Table~\ref{table4} together with the corrected integrated intensities and fluxes (in erg~cm$^{-2}$~s$^{-1}$~sr$^{-1}$) obtained using the [c$^{3}$/(2\,k\,$\nu^3$)] conversion factor. The integrated intensity ratios strongly depend on the filling factor correction, as very different corrections affect both lines. This new set of results indicates that the [OI]$_{\rm 63 \mu m}$/CO (16--15) integrated intensity ratio is significantly lower in the HH377 bow-shock than in the jet and arguably in the outflow cavity, as well. We note that valuable information would be gained from a fully sampled map of this region with SOFIA-GREAT. This would allow us to study the extension of the emission spectral component per spectral component, although the spatial resolution of SOFIA might not suffice to reach constraining conclusions.

\subsection{Abundances}
\label{sub:abundances}

We finally computed the \ion{O}{I} column density. We assumed optically thin emission and LTE conditions. Both assumptions are discussed at the end of this section. Respectively denoting with $h$, $k$, $c$, $\nu$, $E_{\rm u}$, $A_{\rm ul}$, and $g_{\rm u}$ Planck's constant, Boltzmann's constant, the speed of light, the line frequency, upper level energy, Einstein's coefficient, and the upper level statistical weight of the observed transition (see Table~\ref{table2} for these last parameters), we hence used the following formula:
\begin{equation}
N_{\rm OI}{\rm(cm^{-2})} = \frac{8\pi k \nu^2}{hc^3} \frac{10^5}{A_{\rm ul}} \frac{Q_{\rm Tex} exp(E_{\rm u}/kT_{\rm ex})} {g_{\rm u}} \int T_{\rm MB} d \varv\, \rm{(K~km~s^{-1})}
\label{equation1}
\end{equation}

In their analysis of CO lines excitation, L15 identified a low-excitation and a high-excitation component in the jet and the outflow cavity, and a high-excitation component was associated with HH377. The main properties inferred by L15 for each of these components are listed in Table~\ref{table5}. In order to calculate a column density and column density ratio for \ion{O}{I}, it is necessary to understand the component of origin for the [OI]$_{\rm 63 \mu m}$ emission. Unfortunately, the three kinematic components (jet, HH377, and outflow cavity) are all seen in low- and high-excitation lines (CO 2--1 and 16--15, respectively), and the upper level of our observed \ion{O}{I} transition has spectroscopic parameters ($E_{\rm up} = 227.7$~K, $A_{\rm ij} = 8.9 \times 10^{-5}$~s$^{-1}$) similar to those of the CO $J_{\rm up} = 9$ transition ($E_{\rm up} = 248.9$~K, $A_{\rm ij} = 7.3 \times 10^{-5}$~s$^{-1}$), where the transition from low- to high-excitation occurs in the CO rotational diagram of the jet and the outflow cavity. We consequently calculated the column density of \ion{O}{I} assuming that its emission originates from both the low- and high-excitation components for the jet and the outflow cavity. Our results are given in Table~\ref{table5}.

\begin{table*}
\caption{\ion{O}{I} abundance calculations. The ($N$(CO), $T_{\rm kin}$, $n$) values are those determined by L15 for the \lq low-' or \lq high'-excitation component of the CO  observations (see text). The uncertainty on $N$(\ion{O}{I}) accounts for the uncertainty on the $T_{\rm ex}$ value, integrated intensity value, and filling factor correction.}            
\label{table5}      
\centering                          
\begin{tabular}{l l c c c}     
\hline\hline           
& Component    &   Jet & Bow-shock & outflow cavity \\

\hline





\multirow{5}{*}{low-excitation assumption}

& $N$(CO) (10$^{16}$~cm$^{-2}$) & 9.0 & -- & 70.0\\

& $T_{\rm kin}$ (K) & 80--100 & -- & 55--85 \\

& $n$  (cm$^{-3}$) & (0.5--1)$\times 10^{5}$ & -- & (1--8)$\times 10^{5}$ \\

& $N$(\ion{O}{I}) (10$^{16}$~cm$^{-2}$) & 24.6$\pm$8.5 & -- & > 24.8 \\

& $N$(\ion{O}{I}) / $N$(CO) & 2.7$\pm$0.9 & -- & >0.4 \\

\hline

\multirow{5}{*}{high-excitation assumption}

& $N$(CO) (10$^{16}$~cm$^{-2}$) & 1.5 & 10.0 & 6.0\\

& $T_{\rm kin}$ (K) & 400--750 & 400--500 & 500--1500 \\

& $n$  (cm$^{-3}$) & (0.5--1)$\times 10^{6}$ & (1.0--2.0)$\times 10^{6}$ & (1--5)$\times 10^{6}$ \\

& $N$(\ion{O}{I}) (10$^{16}$~cm$^{-2}$) & 4.0$\pm$1.0 & 2.1$\pm$0.6 & > 4.9\\

& $N$(\ion{O}{I}) / $N$(CO) & 2.7$\pm$0.6 & 0.2$\pm$0.1 & > 0.8 \\

\hline\hline
\end{tabular}
\end{table*}

These calculations show that the \ion{O}{I} column density is higher in the outflow cavity than in the jet, itself presenting a higher \ion{O}{I} column density than the terminal shock of the Cep E outflow. In terms of  \ion{O}{I} quantity over the entire outflow, this effect is accentuated by the larger extension of the outflow cavity with respect to the jet, let alone to the HH377 region. The abundance ratio of \ion{O}{I} to CO is also lowest in the terminal shock region (about 0.2, see Table~\ref{table5}). Finally, our strongest conclusion from Table~\ref{table5} is that the jet component is atomic ($N$(\ion{O}{I}) / $N$(CO)$\sim$2.7) regardless of the assumption that we made on the origin of the [OI]$_{\rm 63 \mu m}$ emission. 

The LTE assumption is only validated in the jet, bow-shock, and outflow cavity under the high-excitation assumption by the local density values constrained by L15. These are indeed greater than the critical density of the [OI]$_{\rm 63 \mu m}$ line, about 5$\times 10^5$~cm$^{-3}$ (see Table~\ref{table5}). The LTE approximation is not so well validated under the low-excitation assumption, where the local densities are lower. This means that the \ion{O}{I} column densities obtained in this way are most likely lower limits to the column densities and not definitive values. This does not crucially modify the conclusions of the above paragraph. 

The optically thin assumption could also be questioned. We adopted a line width of 15~km~s$^{-1}$ for the jet component, consistent with our observations and those of L15. Using RADEX online (\citealt{Vandertak07}), we found that the [OI]$_{\rm 63 \mu m}$ line only becomes thick if $N$(\ion{O}{I}) exceeds 3.1$\times 10^{18}$ and 6.3$\times 10^{18}$~cm$^{-2}$ in the low- and high-excitation conditions, respectively, that is, a factor 10 to 100 times higher than the \ion{O}{I} column densities we found.

The adequate remedy to improve our column density measurement would be to map the region also in the $^3$P$_0 \rightarrow$ $^3$P$_1$ transition of \ion{O}{I} (at 145~$\mu$m). This transition is indeed likely to be optically thin, and we could hence apply an LVG treatment to these two lines. We would thus overcome the limitations intrinsic to the LTE approach and be able to study the origin of the [OI]$_{\rm 63 \mu m}$ emission (from the low- or high-excitation CO component). SOFIA/upGREAT will allow this observation in a near future. The [OI]$_{\rm 63 \mu m}$/[OI]$_{\rm 145 \mu m}$ line ratio predicted by \citet{Nisini15} would make this detection difficult. They are in the 25--50 range (depending on the kinetic temperature and main collision partner, see their Figure 10) for the values of the local density of the high-excitation component inferred by L15. On the other hand, the ratios we found with RADEX online are significantly lower (of the order of 1.5--3), and the $^3$P$_0 \rightarrow$ $^3$P$_1$ transition of \ion{O}{I} at 145~$\mu$m has been successfully detected by \textit{Herschel}-PACS.

\section{Nature of shocks in Cepheus E}
\label{sec:nature}

\subsection{Shock modelling of the jet component}
\label{sub:jet}

In L15, we successfully reproduced the CO emission in the high-excitation component of the jet by means of shock models computed with the Paris-Durham code. We compared the observations to a grid of such models and found several solutions depending on filling factor assumptions of this high-excitation component. We recall the solutions here and compare them to our [OI]$_{\rm 63 \mu m}$ observations in an attempt to lift the degeneracy of the solution by fitting a maximum number of observational data.

Without a filling factor correction in addition to the correction that is based on the interferometric observation of CO (2--1) emission, six models were found to provide a satisfying fit to the CO observations of the jet. All of them were young non-stationary (\citealt{Lesaffre041, Lesaffre042}) shock models with the following parameters: pre-shock density $n_{\rm H}$ between 5$\times 10^4$ or 10$^5$~cm$^{-3}$, magnetic field parameter $b = 1$ ($B$($\mu$G) = $b[n_{\rm H}\rm{(cm^{-3})}]^{1/2}$, where $B$ is the transverse magnetic field strength), shock velocity $\varv_{\rm s}$ between 20 and 30~km~s$^{-1}$, and age between 225 and 435~yr. When we applied an additional filling factor of 0.25 to the CO data (effectively multiplying the integrated intensities by a factor 4) to account for the possibility that the high-excitation component is less extended than the bulk of the CO emission, six older models fit the data, with $n_{\rm H} = 10^5$~cm$^{-3}$, $b$ = 1, $\varv_{\rm s}$ = 25--30~km~s$^{-1}$, and age between 595 and 740~yr. These models and their associated [OI]$_{\rm 63 \mu m}$ flux are given in Table~\ref{table6}.

\begin{table}[h]
\small
\caption{Predictions for the [OI]$_{\rm 63 \mu m}$ line emission from the models that were found to be a satisfying fit to the high-excitation CO emission from the jet by L15. For all shock models, $b$ = 1 (see text). The first block of models was obtained for a filling factor of 1, the second for a filling factor of 0.25.}            
\label{table6}      
\centering                          
\begin{tabular}{c  c }        
\hline        
\hline              
shock parameters & [OI]$_{\rm 63 \mu m}$ emission \\
($n_{\rm H}$, $\varv_{\rm s}$, age) & (erg cm$^{-2}$ s$^{-1}$ sr$^{-1}$) \\
\hline
5$\times 10^4$~cm$^{-3}$, 25--30~km~s$^{-1}$, 250--315 yr & [2.9$\times 10^{-6}$--2.8$\times 10^{-4}$]\\
10$^5$~cm$^{-3}$, 20~km~s$^{-1}$, 225--280 yr & [6.3$\times 10^{-6}$--6.6$\times 10^{-6}$]\\
\hline
10$^5$~cm$^{-3}$, 25--30~km~s$^{-1}$, 595--740 yr & [4.5$\times 10^{-6}$--6.3$\times 10^{-6}$]\\
\hline           
\hline           
\end{tabular}
\end{table}

The first conclusion is that the predicted value is very dependent on the shock parameters, as our results are spread over two orders of magnitude, between 2.9$\times 10^{-6}$ and 2.8$\times 10^{-4}$~erg~cm$^{-2}$~s$^{-1}$~sr$^{-1}$. The most important parameter to explain this spread is the age of the shock, which determines the strength of the J-type contribution, and that of the [OI]$_{\rm 63 \mu m}$ line emission in our non-stationary models. The observed value in the jet component is $2.3\pm0.5 \times 10^{-3}$~erg~cm$^{-2}$~s$^{-1}$~sr$^{-1}$. If an additional filling factor of 0.25 is applied, similarly to what was considered for CO by L15, these values should be multiplied by 4. In any case, we find that none of our shock models approaches the observed value, as they predict at least a factor 10 less \ion{O}{I} emission than observed. The first explanation is that the bulk of the [OI]$_{\rm 63 \mu m}$ line emission comes from the (not shock-modelled) low-excitation component instead of the high-excitation component.

However, if the [OI]$_{\rm 63 \mu m}$ line emission comes from the high-excitation component, a number of explanations could account of the discrepancy between observations and shock models. A first explanation is that our model is inaccurate regarding the oxygen chemistry. The main oxygen carriers are CO, H$_2$O, OH, O$_2$, and \ion{O}{I}. The models that best fit the CO emission predict that the local fractional abundance of both CO and H$_2$O are about 10$^{-4}$ in the post-shock region and that the local fractional abundance of O$_2$ is greater than that of \ion{O}{I}. The high value of the local fractional abundance of O$_2$ is a well-documented problem in our simulation and in chemical models in general (see e.g. \citealt{Viti01} and \citealt{Hollenbach09} for a discussion of this specific aspect). Meaningful constraints on the oxygen chemistry could then be obtained through the observation of H$_2$O and O$_2$. Unfortunately, no velocity-resolved observations of water have been performed in Cep E by \textit{Herschel}, and no instrument can allow them at the moment. O$_2$ is best observed from space observatories (the $N_{\rm J} = 3_3 - 1_2$ transition at 487.249~GHz is close to an atmospheric absorption feature), but it has not been observed in Cep E, and its symmetry makes its detection difficult from the outset.

Another possibility is that the shock that we observed is illuminated by the protostar that drives the outflow. In such irradiated shocks (e.g. \citealt{Lesaffre13, Melnick15}), both H$_2$O and CO can be photo-dissociated, increasing the gas-phase abundance of \ion{O}{I}. This effect is not taken into account in the Paris-Durham code. However, the possibility that shocks at position BII could be affected by the distant protostar is relatively low: the BII and mm position are 8.4$\times 10^{-2}$ pc away. The Cep E-mm protostar has a luminosity of about 100~$L_\odot$, and a T-Tauri star typically emits between 10$^{-1.5}$ and 10$^{-3.5}$ of its total luminosity in the UV range between 3200 and 5200 \AA~\citep{Gullbring98}. Conservatively assuming that Cep E-mm is radiating 10\% of its total luminosity in this UV range and taking dilution into account, we found that the radiation field in BII is about 4.8~$G_{\rm 0}$ in units of Draine's (\citealt{Draine78}) or Mathis's (\citealt{Mathis83}) interstellar radiation field (ISRF). This value has then to be corrected for the absorption by dust grains. We assumed an average local density in the outflow cavity of 10$^{-4}$~cm$^{-3}$, which translates into a column density of $N_{\rm{H}} \simeq 2.6\times10^{21}$~cm$^{-2}$ and into a visual extinction of $A_{\rm{v}} \simeq 1.4$ mag (\citealt{Lefloch96} provided local estimates for this value: 61 mag around the driving protostar, and 3.2--3.4 mag in the H$_2$ knots along the flow). Using the exponential attenuation approximation $I = I_{\rm 0} \times e^{-\tau}$, where $\tau = N_{\rm H} \times \sigma$, and $\sigma$ is the effective attenuation cross section from \citet{Draine96}, we found that the radiation field really felt in BII is about 2.6$\times 10^{-3} G_{\rm 0}$ in Draine's or Mathis's ISRF units.

Another explanation is that several shock structures are present in our beam. In this case, the missing \ion{O}{I} emission could be accounted for by the presence of another shock structure. This additional shock layer must present different physical characteristics, as simply combining shock layers similar to those we used would simultaneously increase the [OI]$_{\rm 63 \mu m}$ emission insufficiently and increase the CO emission too much, producing another type of discrepancy between observations and models. One such shock structure could be a stationary dissociative J-type shock. In particular, a J-type shock with $n_{\rm H} = 10^4$~cm$^{-3}$, magnetic field parameter $b = 0.1$, and shock velocity $\varv_{\rm s} \geq 30$~km~s$^{-1}$ \citep{Flower15} could be combined with our current best-fit solution, providing a satisfying fit for the \ion{O}{I} and CO emission. 

The additional shock structure could also be a shock with a radiative precursor, or in other words, a shock structure whose high temperatures generate FUV-illumination that affects both pre- and post-shock regions \citep{Hollenbach89}. Figure 7 of \citealt{Hollenbach89} shows that these types of shocks (especially with a pre-shock density $n_{\rm H} = 10^4$~cm$^{-3}$) could reproduce the observed levels of the [OI]$_{\rm 63 \mu m}$ line emission. However, these shocks also produce CO emission (see Figure 6 of \citealt{Hollenbach89}): combining this solution with our models would perhaps prove successful at fitting the [OI]$_{\rm 63 \mu m}$ emission, but would generate a discrepancy in the CO comparisons. On the other hand, it is possible that the CO and [OI]$_{\rm 63 \mu m}$ line emission could both arise from the same shock structure with a radiative precursor. 

All these explanations (including the possibility that the [OI]$_{\rm 63 \mu m}$ emission comes from the low-excitation component) deserve to be explored in more detail by comparing a maximum of observables with the models. In this respect, the important role of SOFIA/upGREAT can only be stressed, as the observation of the \ion{O}{I} $^3$P$_0 \rightarrow$ $^3$P$_1$ line at 145~$\mu$m and the $^2$P$_{3/2} \rightarrow$ $^2$P$_{1/2}$ \ion{C}{II} line at 158~$\mu$m will allow placing more meaningful constraints on our observations and existing models. 

\subsection{Comparison with other star-forming regions}
\label{sub:cwsfr}

   \begin{figure}[t]
   \centering
   \includegraphics[width=8cm]{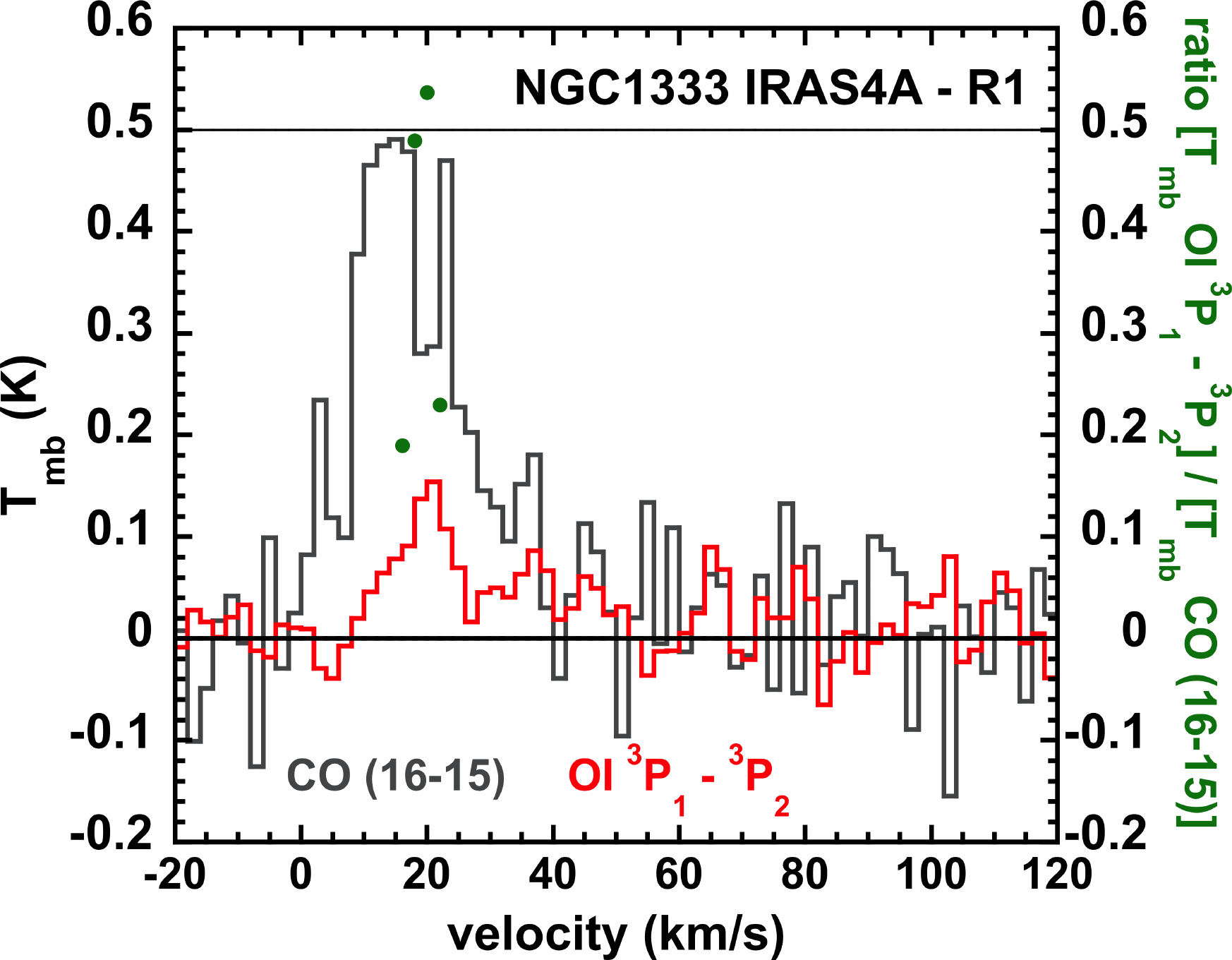}
   \includegraphics[width=8cm]{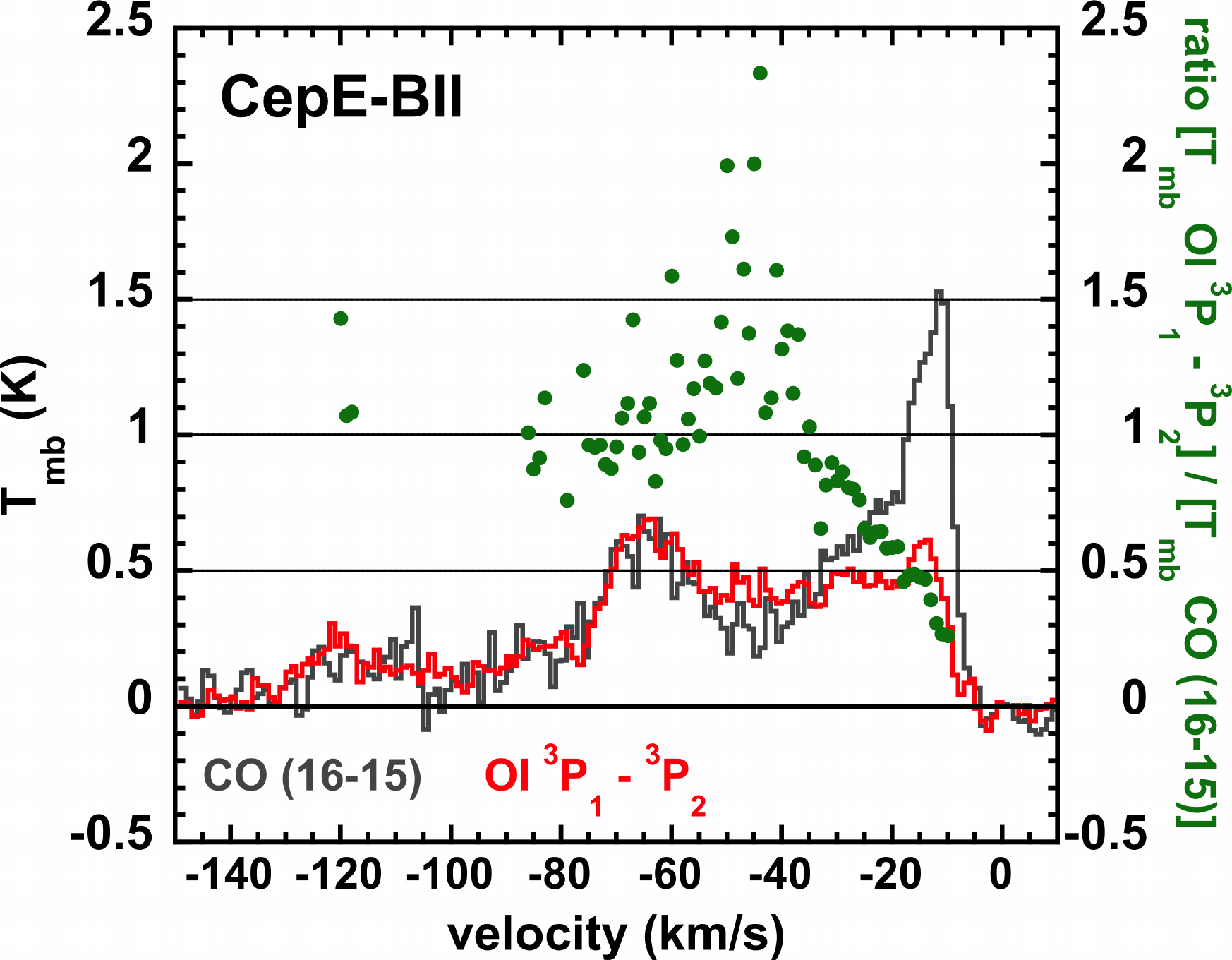}
   \includegraphics[width=8cm]{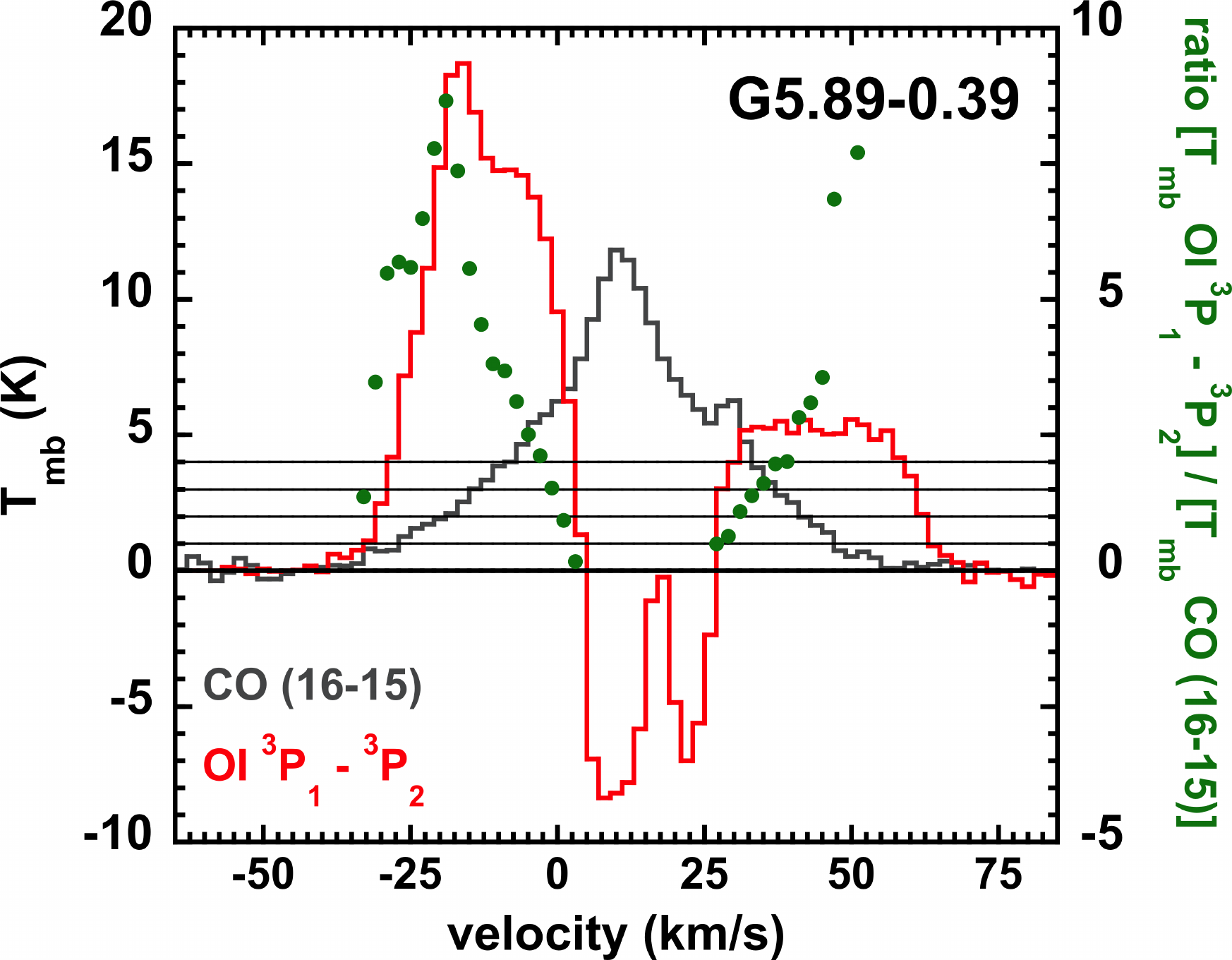}
      \caption{[OI]$_{\rm 63 \mu m}$ (red line) and CO (16--15) (grey line) spectra, and \{$T_{\rm mb}$ [OI$_{\rm 63 \mu m}$]\} / \{$T_{\rm mb}$ [CO (16-15)]\} ratio (green points) obtained at nominal spatial resolutions, channel by channel, in NGC1333-IRAS4A RI (upper panel, from Kristensen et al., submitted), Cep E-BII (middle panel), the central position of the G5.89--0.39 massive star-forming region (from \citealt{Leurini15}). The line ratio is only shown for channels where both lines have been detected above the 3$\sigma$ threshold. Horizontal black lines at 0.5, 1, 1.5, and 2 values are shown in relation with this ratio as a guide for the eye in both panels.}
         \label{figure3}
   \end{figure}

We recall that the BII spectrum has been obtained at a position that is distant from the protostar ($r$ = 0.084~pc), which is located at $d$ =730 pc from the Earth and has a luminosity $L = 100~L_\odot$. To compare our observations with other regions where [OI]$_{\rm 63 \mu m}$ and CO (16--15) have been observed in similar conditions, we plot the two spectra and their channel-by-channel ratio in the upper panel of Fig.~\ref{figure3}. In this panel, the ratio is shown as green points for every channel where the two lines have been detected above the 3$\sigma$ level. We extracted the same spectra and ratio from two other star-forming regions. The first is the R1 shock spot of the NGC1333-IRAS4A bipolar outflow, which is driven by a low-mass protostar (for which $r$ = 0.011~pc, $d$ = 235~pc, $L$ = 9.1$L_\odot$; \citealt{Kristensen16}). In this region, the [OI]$_{\rm 63 \mu m}$ filling factor is of the order of unity, as shown by the half-maximum contour of Figure 2 of \citet{Nisini15}. The second region is the central position of G5.89--0.39, a complex high-mass star-forming region with several outflows and spectacular [OI]$_{\rm 63 \mu m}$ emission (for which $r$ = 0~pc, $d$ = 1.28~kpc, $L$ = 1.3$\times 10^5 L_\odot$; \citealt{Leurini15}). In this region, the [OI]$_{\rm 63 \mu m}$ filling factor is also of the order of unity given the size of the emiting region. The comparison between the three datasets is shown in Fig.~\ref{figure3}. 

The line temperature ratio is very different in the three cases. In NGC1333-IRAS4A, the [OI]$_{\rm 63 \mu m}$/CO (16--15) temperature ratio does not exceed the value of 0.6. In this case, only four points are shown that correspond to the four velocity channels where [OI]$_{\rm 63 \mu m}$ emission is detected above 3$\sigma$. In Cep E-BII, the great majority of the ratio values is between 0.5 and 1.5, with lower values found in the outflow cavity component near the systemic velocity. Finally, in G5.89--0.39,  the ratio is lower than 2 only at velocities close to the systemic velocity (between -5 and 40~km~s$^{-1}$ for a $\varv_{\rm lsr} \sim$10~km~s$^{-1}$). Given that the [OI]$_{\rm 63 \mu m}$ emission is self-absorbed at these velocities \citep{Leurini15}, these points do not reflect the local shock conditions. Moreover, it is possible that self-absorption affects the emission for velocities between -20 to 50~km~s$^{-1}$. It follows that the ratio is systematically in the 5--20 range. Although not corrected for filling factor effects, this is significantly higher than what is seen in Cep E and NGC1333-IRAS4A. 

Depending on filling factor corrections, this [OI]$_{\rm 63 \mu m}$/CO (16--15) temperature ratio could be reflective of the nature of shocks in the different sources:
\begin{itemize}
\item in the low-mass bipolar outflow NGC1333-IRAS4A the observed line width seems to suggest that the shock propagates at a moderate velocity and is hence likely molecular. Additionally, the region is offset from the (low-luminosity) protostar, thus not irradiated by its far-UV-illumination. In such conditions, non-dissociative $C$-, $J$- or $CJ$-type shock models could be used to interpret the observations;
\item in Cep E-BII, we have seen that those models are probably not sufficient to fit the observations in the jet, where additional dissociative shocks or alternative solutions with radiative precursors must be invoked;
\item finally, in G5.89--0.39, where the observations were centred on the very luminous protostar, the very high intensity of [OI]$_{\rm 63 \mu m}$ emission could be a signature of the effect of the far-UV field of the central protostar on the various outflows that propagate in the region, all the more so because the filling-factor-corrected ratio is in the 1--5 range. 
\end{itemize}
This global picture of the [OI]$_{\rm 63 \mu m}$/CO (16--15) interpretation needs to be tested through thorough comparisons between additional observations and various models. In particular, the observation of the $^2$P$_{3/2} \rightarrow$ $^2$P$_{1/2}$ \ion{C}{II} line at 158~$\mu$m with SOFIA/upGREAT, combined with ground-based observations of CO and \ion{C}{I}, will allow us to quantify dissociation effects in the various targeted regions.

\section{Concluding remarks}
\label{sec:conc}

\begin{enumerate}
\item We have used the SOFIA/GREAT instrument to observe the emission of the \ion{O}{I} $^3$P$_1 \rightarrow$ $^3$P$_2$, OH between $^2\Pi_{1/2}$ $J = 3/2$ and $J = 1/2$ at 1837.8~GHz and CO (16--15) lines towards three positions in the Cepheus E outflow (namely Cep E-mm, Cep E BI, and Cep E BII). The CO (16--15) line was detected at all three positions. The [OI]$_{\rm 63 \mu m}$ line was detected in Cep E BI and BII, whereas the OH line was not detected. 
\item We focused our study on the most significant detection in Cep E BII. Three kinematical components were detected in the [OI]$_{\rm 63 \mu m}$ spectrum, which are remarkably similar to those seen in CO (16--15). These are themselves similar to those seen by L15 in other CO transitions, and related to various spatial components: the jet, the HH377 terminal bow-shock, and the cavity walls all contribute to the line emission. This allowed us to estimate \lq most realistic' filling factor values for the [OI]$_{\rm 63 \mu m}$ emission in each of these components based on the observations presented in L15: 0.25$\pm$0.05, 0.55$\pm$0.15, and 1.
\item We used the results of L15 and analysed the [OI]$_{\rm 63 \mu m}$ emission to infer \ion{O}{I} column densities and $N$(\ion{O}{I}) / $N$(CO) ratios in each of these components. The two quantities depend on the filling factor correction and on the characteristics of the region of origin for the [OI]$_{\rm 63 \mu m}$ emission. The \ion{O}{I} column density is higher in the outflow cavity than in the jet, which itself presents a higher \ion{O}{I} column density than the terminal shock. In terms of  \ion{O}{I} quantity over the entire outflow, this effect is accentuated by the larger extension of the outflow cavity with respect to the jet, let alone to the HH377 region. The terminal shock is also the region where the abundance ratio of \ion{O}{I} to CO is the lowest (about 0.2). The jet component is atomic ($N$(\ion{O}{I}) / $N$(CO)$\sim$2.7) regardless of the assumption that we made about the origin of the [OI]$_{\rm 63 \mu m}$ emission. 
\item We compared the observed [OI]$_{\rm 63 \mu m}$ flux in the jet component with shock models from the Paris-Durham code that proved successful to fit the high-excitation emission of the jet component of 10 CO lines. All of these models are at least a factor 10 below the observed [OI]$_{\rm 63 \mu m}$ flux. This discrepancy might indicate that the \ion{O}{I} emission comes from the low-excitation component. This discrepancy could also be a sign that the current chemical modelling of oxygen is not accurate enough, or that the shocks are irradiated by the protostar. Two more compelling explanations exist for this discrepancy: several shock structures (including dissociative $J$-type shocks) could be present in the beam of our observations, or the CO and \ion{O}{I} emission could arise from a $J$-type shock with a radiative precursor. 
\item We compared our observations in the Cep E-BII region with similar observations of other star-forming regions: one shock position in a bipolar outflow driven by a low-mass protostar, NGC1333-IRAS4A, and one position centred on a massive protostar at the centre of several outflow structures in G5.89--0.39. The trend observed in the channel-by-channel [OI]$_{\rm 63 \mu m}$/CO (16--15)  temperature ratio (which increases with the mass of the central protostar) could be a signature of various types of shocks propagating in these regions.
\item Most of the limitations that we have encountered in the present study will find their remedy in the next round of SOFIA/upGREAT observations that could allow us to map the entire outflow in \ion{O}{I} lines (the $^3$P$_1 \rightarrow$ $^3$P$_2$ line at 63~$\mu$m and the $^3$P$_0 \rightarrow$ $^3$P$_1$ line at 145~$\mu$m). This will allow us to improve the accuracy of all our measurements. We would then better be able to constrain the size of emitting regions, the excitation conditions for \ion{O}{I}, and also conclude more precisely about energetic impacts by extending our conclusions to the full outflow. Finally, the observation of the $^2$P$_{3/2} \rightarrow$ $^2$P$_{1/2}$ \ion{C}{II} line at 158~$\mu$m with SOFIA/upGREAT is also an interesting prospect, as it will allow us to more tightly constrain the types of shocks that can fit the data.
\end{enumerate}

\begin{acknowledgements}
We thank an anonymous referee for meaningful comments that helped to improve the clarity and quality of this study. We thank the SOFIA operations and the GREAT instrument teams, whose support has been essential for the GREAT accomplishments, and the DSI telescope engineering team. Based [in part] on observations made with the NASA/DLR Stratospheric Observatory for Infrared Astronomy. SOFIA Science Mission Operations are conducted jointly by the Universities Space Research Association, Inc., under NASA contract NAS2-97001, and the Deutsches SOFIA Institut, under DLR contract 50 OK 0901. B. Lefloch acknowledges support from a grant from LabeX Osug\@2020 (Investissements d'avenir - ANR10LABX56). This work was supported by the French program \lq Physique et Chimie du Milieu Interstellaire' (PCMI) funded by the Conseil National de la Recherche Scientifique (CNRS) and Centre National d'\'Etudes Spatiales (CNES).
\end{acknowledgements}

%
%

\bibliographystyle{aa}
\bibliography{biblio}   

\section{\textit{Herschel}-PACS footprint of the [OI]$_{\rm 63 \mu m}$ emission in the southern lobe of the Cep E outflow}
\label{sec:PACS}

\begin{figure}[b]
   \centering
   \includegraphics[width=9cm]{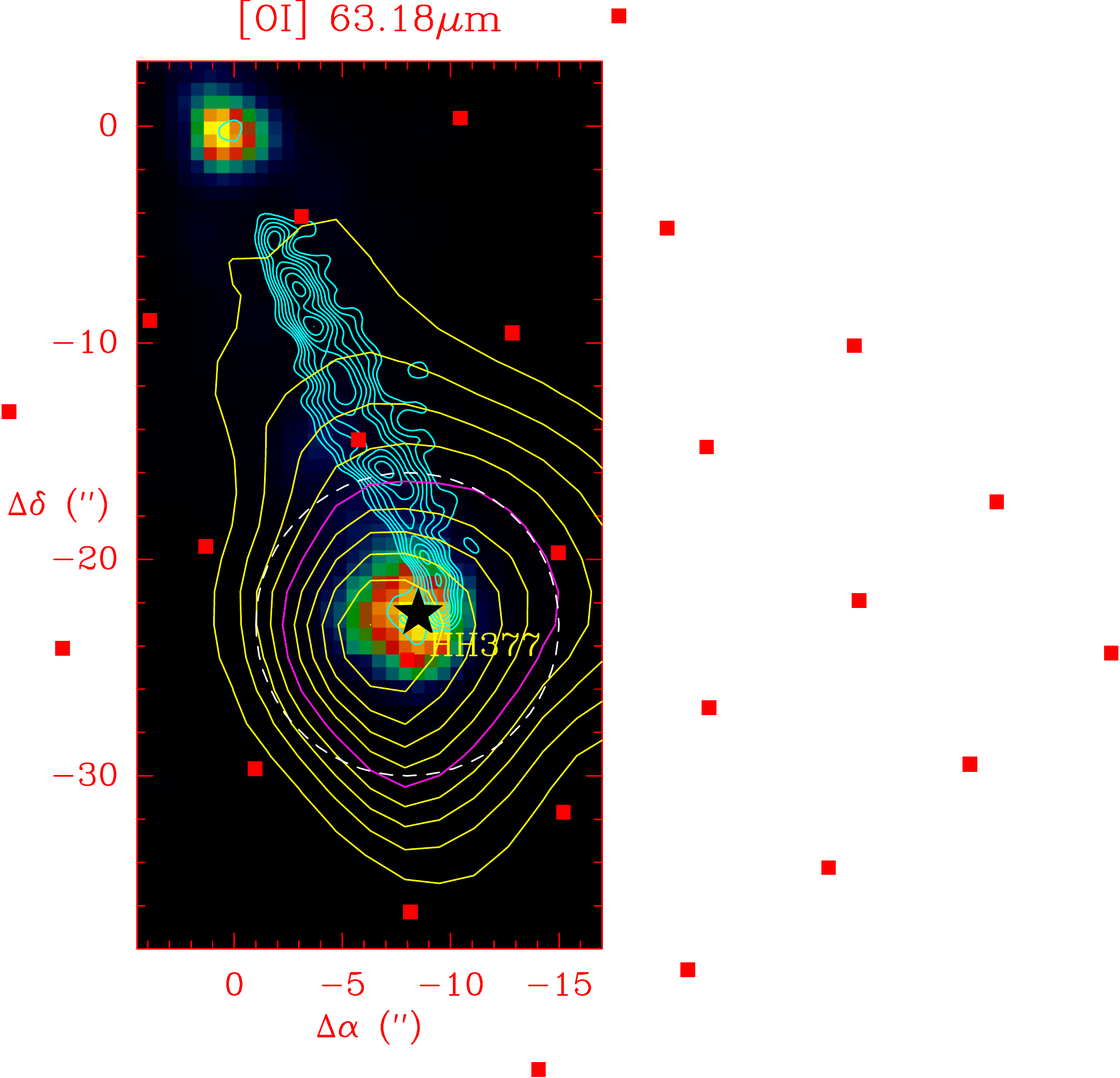}
      \caption{\textit{Spitzer}-IRAC band-one (3.6 $\mu$m) image of the southern lobe of the Cep E outflow (in colours), retrieved from the \textit{Spitzer} archive, overlaid with the CO (2--1) cyan contours of the jet (PdBI data, from L15), and with the [OI]$_{\rm 63 \mu m}$ emission (yellow contours) seen by \textit{Herschel}-PACS. The HPBW contour is plotted in purple and roughly corresponds to a 14$''$ circle shown as a white dashed contour. The red squares indicate the position of the spaxels, and the black star indicates the position of HH377, the terminal bow-shock.}
         \label{figurea1}
   \end{figure}

In this Appendix, we show a footprint of the [OI]$_{\rm 63 \mu m}$ emission obtained in the southern lobe of the Cep E outflow (Figure~\ref{figurea1} with the PACS receiver \citep{Poglitsch10} onboard the \textit{Herschel} space observatory \citep{Pilbratt10}. The image corresponds to the ObsId 1342262543 and was obtained in line spectroscopy mode. It was created using HIPE version 14.2.1 \citep{Ott10}. We solely use the image for the purpose of discussing our filling factor assumptions. The more thorough presentation and deeper analysis of all PACS data (including water line emission) will be the subject of a forthcoming publication.
 
The half-power contour shown in purple in Figure~\ref{figurea1} has a typical diameter of 14$''$ (also see the dashed white contour in the figure). This indicates a typical size of 10$''$ for the emitting region, after rudimentary deconvolution from the under-sampled PACS footprint, assuming each spaxel has a Gaussian size of 9$\farcs$4. This value is indicative of the emission size that is most likely dominated at this position by the outflow cavity.


\end{document}